\documentclass[twocolumn,pra,aps]{revtex4}
\begin{document}
\title{Locality and Nonlocality in  Quantum Pure-state Identification Problems}
\author{Y. Ishida, T. Hashimoto, M. Horibe, and A. Hayashi}
\affiliation{Department of Applied Physics\\
           University of Fukui, Fukui 910-8507, Japan}

\begin{abstract}
Suppose we want to identify an input state with one of two unknown reference states, where the input 
state is guaranteed to be equal to one of the reference states. We assume that no classical knowledge 
of the reference states is given, but a certain number of copies of them are available instead. 
Two reference states are independently and randomly chosen from the state space in a unitary invariant 
way. This is called the quantum state identification problem, and the task is to optimize the mean 
identification success probability. 
In this paper, we consider the case where each reference state is pure and bipartite, and  
generally entangled. The question is whether the maximum mean identification success probability can 
be attained by means of a local operations and classical communication (LOCC) measurement scheme. 
Two types of identification problems are considered when a single copy of each reference state is available. 
We show that a LOCC scheme attains the globally achievable identification probability in the 
minimum-error identification problem. In the unambiguous identification problem, however, the maximal 
success probability by means of LOCC is shown to be less than the globally achievable identification 
probability. 
\end{abstract}

\maketitle

\newcommand{\ket}[1]{|\,#1\,\rangle}
\newcommand{\bra}[1]{\langle\,#1\,|}
\newcommand{\braket}[2]{\langle\,#1\,|\,#2\,\rangle}
\newcommand{\bold}[1]{\mbox{\boldmath $#1$}}
\newcommand{\sbold}[1]{\mbox{\boldmath ${\scriptstyle #1}$}}
\newcommand{\tr}[1]{{\rm tr}\!\left[#1\right]}
\newcommand{\trm}{{\rm tr}}
\newcommand{\BC}{{\bold{C}}}
\newcommand{\CS}{{\cal S}}
\newcommand{\CA}{{\cal A}}
\newcommand{\CM}{{\cal M}}

%%%%%%%%%%%%%%%%%%%%%%%%%%%%%%%%%%%%%%%%%%%%%%%%%%%%%%%%%%
%%%%%%%%%%%%%%%%%%%%%%%%%%%%%%%%%%%%%%%            Introduction
%%%%%%%%%%%%%%%%%%%%%%%%%%%%%%%%%%%%%%%%%%%%%%%%%%%%%%%%%%
\section{Introduction}
It is an extremely nontrivial problem to distinguish different states of a 
quantum system by measurement \cite{Helstrom76,Holevo82,Ivanovic87,Dieks88,Peres88}. 
First of all, this is because of the statistical 
nature of quantum measurement, which destroys the state of the system 
to be measured and one cannot clone an unknown quantum state \cite{Wootters82}. 
Another relevant issue is nonlocality of quantum mechanics. When the system to be measured is a 
composite, we can generally obtain more information about  
the system by a global measurement on the whole system than by a combination of local 
measurements on its subsystems \cite{Peres91,Ban97,Sasaki98,Bennett99,Eldar01,Koashi07}.

Let us focus on the problem of distinguishing two pure states of a composite 
system which is shared by two parties. It is a fundamental 
question of quantum-information theory whether the optimal discrimination  
can be obtained by means of a local operations and classical 
communication (LOCC) scheme for the two parties.

Walgate {\it et al}. \cite{Walgate00} showed that any two mutually orthogonal pure states can be 
perfectly distinguished by LOCC. This is rather surprising 
since their result holds regardless of entanglement of the states. 
It has also been shown that any two generally nonorthogonal pure states 
can be optimally discriminated by LOCC: the optimal 
success probability of discrimination by a global measurement can be 
attained by a LOCC protocol. This was shown for two types of  
discrimination problems: the discrimination problem with minimum error \cite{Virmani01} 
where an erroneous guess is allowed, and the unambiguous discrimination 
problem \cite{Ji05,Chen01,Chen02} where no error is allowed but an inconclusive 
result can be produced.  These results can be interpreted as meaning that there is no 
nonlocality in the discrimination of two pure states.

We can consider a different setting for the discrimination problem of two pure 
states. In the usual setting, it is assumed that perfect classical knowledge 
of the two states to be discriminated ($\rho_1$ and $\rho_2$) is given. 
The measurement scheme for the optimal discrimination naturally depends on 
the classical knowledge of the states.  
Instead, let us assume that there is no classical knowledge of the states 
$\rho_1$ and $\rho_2$, but a certain number ($N$) of their copies 
are available as reference states. 
One's task is to correctly identify a given input state $\rho$ with one of 
the reference states $\rho_1$ and $\rho_2$ by means of a measurement on the 
whole state $\rho \otimes \rho_1^{\otimes N} \otimes \rho_2^{\otimes N}$.
When the number of copies $N$ is infinite, the problem is reduced to quantum 
state discrimination. This is because we can always obtain complete classical 
knowledge of a quantum state if we have infinitely many copies of 
the state. 
We call this problem ``quantum state identification.'' The optimal 
success probability by a global measurement scheme has been determined for 
the minimum-error \cite{Hayashi05} and 
unambiguous \cite{Bergou05, Hayashi06} identification problems. 

In this paper, we investigate the pure-state identification 
problem of $N=1$ where the two reference pure states $\rho_1$ and $\rho_2$ are bipartite. 
The input state $\rho$ given to Alice and Bob is guaranteed to be one of the 
reference states $\rho_1$ and $\rho_2$ with prior probabilities  
$\eta_1$ and $\eta_2$. The two reference states are assumed to be independently and randomly 
distributed on the pure-state space in a unitary invariant way. 
Each reference state generated in this way is generally entangled. 
Two types of identification problems will be considered: identification with minimum error and unambiguous identification where no error is allowed.  
For the minimum-error identification, we will demonstrate that Alice and Bob can identify 
the input state by means of a LOCC protocol with the success probability given by the optimal 
global identification scheme. In the case of the unambiguous identification problem, the maximal 
success probability by means of LOCC will shown to be less than the globally achievable identification 
probability.   

%\vfill
\newpage

%%%%%%%%%%%%%%%%%%%%%%%%%%%%%%%%%%%%%%%%%%%%%%%%%%%%%%%%%%%%%%%%%%%%%%%%%%%%%55
%%%%%%%%%%%%%%%%%%%%%%%%%%%%%%%%%%%%%%%%%%%%%%%%%%%%%%% Minimum-error identification without LOCC conditions
%%%%%%%%%%%%%%%%%%%%%%%%%%%%%%%%%%%%%%%%%%%%%%%%%%%%%%%%%%%%%%%%%%%%%%%%%%%%%%%%%
\section{Minimum-error identification without LOCC conditions}
In this section, we will precisely formulate the minimum-error pure-state identification 
problem and derive the maximum mean success probability without the LOCC conditions 
for the case of $N=1$ and an arbitrary prior occurrence probability of 
the reference states. 
In the case of a single-qubit system, the problem has been solved by Bergou {\it et al}.\cite{Bergou06}. 
For the case of general $N$ but with equal prior 
occurrence probabilities, see Ref. \cite{Hayashi05}. 

We have three quantum systems numbered 0, 1, and 2, each on a $d$-dimensional 
space $\BC^{d}$.  The input pure state $\rho=\ket{\phi}\bra{\phi}$ is prepared in system 0 and 
the two reference pure states $\rho_1=\ket{\phi_1}\bra{\phi_1}$ and 
$\rho_2=\ket{\phi_2}\bra{\phi_2}$ in systems 1 and 2, respectively. 
The space that an operator acts on is specified by a number in parentheses. 
For example, $\rho(0)$ is a density operator on system 0. 
The input state $\rho$ is promised to be one of the reference states $\rho_1$ and $\rho_2$ 
with prior probabilities $\{\eta_1, \eta_2\}$. 
The two reference states are independently and randomly chosen from the state space $\BC^{d}$ in  
a unitary invariant way. More precisely, the distribution is assumed uniform 
on the $(2d-1)$-dimensional unit hypersphere of $2d$ real variables 
$\{{\rm Re} c_i, {\rm Im} c_i\}_{i=0}^{d-1}$, 
where $c_i$ are the expansion coefficients of the state 
in terms of an orthonormal base $\{ \ket{i} \}_{i=0}^{d-1}$. The distribution 
does not depend on a particular choice of the base.

Our task is to correctly identify the input state with one of the reference states 
$\rho_\mu (\mu=1,2)$ by measuring the whole system $0 \otimes 1 \otimes 2$. 
We denote the corresponding positive-operator-valued measure (POVM) elements by $E_\mu (\mu=1,2)$. 
The mean identification success probability is then given by 
\begin{eqnarray}
  p(d) = \sum_{\mu=1,2} \eta_\mu \Big<
           \tr{E_\mu \rho_\mu(0)\rho_1(1)\rho_2(2)} \Big>,
\end{eqnarray}
where the symbol $< \cdots  >$ represents the average over the reference states $\rho_1$ and 
$\rho_2$. Note that the POVM $E_\mu$ is independent of $\rho_1$ and $\rho_2$, since 
we have no classical knowledge of the reference states.  

The average over the reference states can be readily performed by using the formula \cite{Hayashi04}:
\begin{eqnarray}
  < \rho^{\otimes n} > = \frac{\CS_n}{d_n}, \label{formula}
\end{eqnarray}
where $\CS_n$ is the projector on to the totally symmetric subspace of $(\BC^d)^{\otimes n}$ 
and $d_n$ is its dimension given by $d_n={}_{n+d-1}C_{d-1}$.
Using $E_2=1-E_1$, the mean success probability to be maximized is written as  
\begin{eqnarray}
  p(d) = \eta_2 + \frac{1}{d_1 d_2} \tr{ E_1( \eta_1\CS(01) - \eta_2\CS(02) ) }, \label{pd}
\end{eqnarray}
where $\CS(01)$ and $\CS(02)$ are the projectors onto the totally symmetric subspaces of 
spaces $0 \otimes 1$ and $0 \otimes 2$, respectively.  

The only restriction on the POVM element $E_1$ is $0 \le E_1 \le 1$. In order to maximize 
the mean success probability Eq.(\ref{pd}), we use the following result which holds for any Hermitian 
operator $\Delta$: 
\begin{eqnarray*}
    && \max_{0 \le E \le 1} \tr{E\Delta} \nonumber \\
    && = \mbox{sum of all positive eigenvalues of $\Delta$},
\end{eqnarray*}
where the maximum is attained when $E$ is the projector $P_+$ onto the subspace $V_+(\Delta)$ 
spanned by all eigenstates of $\Delta$ with a positive eigenvalue. 
Note that it does not matter whether the subspace $V_+(\Delta)$ includes eigenvectors with 
zero eigenvalue.  
In our case, $\Delta$ is defined to be 
\begin{eqnarray}
   \Delta= \eta_1\CS(01) - \eta_2\CS(02). \label{delta}
\end{eqnarray}

Let us decompose the total space into three subspaces according to the symmetry with 
respect to system permutations \cite{Hamermesh62}: 
\begin{eqnarray}
V=\BC^d \otimes \BC^d \otimes \BC^d=V_\CS \oplus V_\CA \oplus V_\CM.
\end{eqnarray}
Here $V_\CS$ is the totally symmetric subspace of dimension $\dim V_\CS \equiv d_3=d(d+1)(d+2)/6$ 
and $V_\CA$ is the totally antisymmetric subspace of dimension $\dim V_\CA =d(d-1)(d-2)/6$. 
The remaining subspace $V_\CM$ is the mixed symmetric subspace of dimension 
$\dim V_\CM =2d(d^2-1)/3$.  
The subspace $V_\CM$ contains the two-dimensional irreducible representation of the symmetric group 
of order 3, $S_3$, with multiplicity $\dim V_\CM /2$. 
We denote projectors onto 
$V_\CS$, $V_\CA$, and $V_\CM$ by $\CS_3$, $\CA_3$ and $\CM_3$, respectively. 

It is clear that $\Delta=\eta_1-\eta_2$ in $V_\CS$ and $\Delta=0$ in $V_\CA$. 
To determine eigenvalues of $\Delta$ in $V_\CM$, it is convenient to introduce two 
operators $D$ and $A$:
\begin{eqnarray}
  &&  D \equiv \CS(01)-\CS(02) = \frac{1}{2} \left( T(01) - T(02) \right),  \label{D} \\
  &&  A \equiv \CS(01)+\CS(02)-1 = \frac{1}{2} \left(  T(01) + T(02) \right). \label{A}
\end{eqnarray}
Here, $T(01)$ is the operator that exchanges systems 0 and 1 and $T(02)$ exchanges 
systems 0 and 2. Calculating $D^2$, we find 
\begin{eqnarray*}
   D^2 &=& \frac{1}{4} \left( 2-T(01)T(02)-T(02)T(01) \right) \\
       &=& \frac{3}{4} \left(1 - \CS_3 -\CA_3 \right) \\
       &=& \frac{3}{4} \CM_3,
\end{eqnarray*}
which implies that eigenvalues of $D$ are $\pm \sqrt{3}/2$ in $V_\CM$ and 0 otherwise. 
It is also easy to show that 
\begin{eqnarray} 
  && DA+AD=0, \label{anti} \\
  && A^2 = 1-D^2. \label{asqure}
\end{eqnarray}
The anticommutability of Eq.(\ref{anti}) implies that if $\ket{+}$ is an eigenstate of $D$ 
with eigenvalue $\sqrt{3}/2$, then $A \ket{+}$ is also an eigenstate of $D$ with eigenvalue 
$-\sqrt{3}/2$. By Eq.(\ref{asqure}), we find that $\ket{-} \equiv 2A\ket{+}$ is correctly 
normalized. Note that the positive and negative eigenvalues of $D$ have the same 
multiplicity. 
Thus we can choose the orthonormal base $\{ \ket{+,k}, \ket{-,k} \}$ 
in $V_\CM$ such that 
\begin{eqnarray*}
  && D\ket{+,k} = + \frac{\sqrt{3}}{2}\ket{+,k}, \\ 
  && D\ket{-,k} = - \frac{\sqrt{3}}{2}\ket{-,k}, \\ 
  && A\ket{+,k} = \frac{1}{2}\ket{-,k}, \\
  && A\ket{-,k} = \frac{1}{2}\ket{+,k},
\end{eqnarray*}
where the index $k$ runs from 1 to $\dim V_\CM/2$.
In this base, $D$ and $A$ are block-diagonalized with respect to $k$ and 
each block has the following $2 \times 2$ matrix representation:
\begin{eqnarray*}
  D = \left( \begin{array}{cc}
                 \frac{\sqrt{3}}{2}  &   0                  \\
                 0                   &  -\frac{\sqrt{3}}{2} \\
             \end{array}
      \right),\ \ 
  A = \left( \begin{array}{cc}
                 0           &  \frac{1}{2}  \\
                 \frac{1}{2} &  0            \\
             \end{array}
      \right).
\end{eqnarray*}

In terms of $D$ and $A$, the operator $\Delta$ is written as 
\begin{eqnarray}
  \Delta = \frac{1}{2} \left( \eta_1-\eta_2 + D + (\eta_1-\eta_2)A \right). \label{deltaDA}
\end{eqnarray}
The operator $\Delta$ is also block-diagonalized with the same $2 \times 2$ matrix representation 
which can be readily diagonalized. Two eigenvalues of $\Delta$ are given by
\begin{eqnarray}
   \lambda_{\pm} = \frac{1}{2} \left( \eta_1-\eta_2 \pm \sqrt{1-\eta_1\eta_2} \right), 
\end{eqnarray}
and we find that  $\lambda_+ \ge 0$ and $\lambda_- \le 0$. 

Now we can calculate the maximum success probability. Let us assume $\eta_1 \ge \eta_2$ 
for the moment. The positive eigenvalues of $\Delta$ are $\eta_1 - \eta_2$ in $V_\CS$ 
with multiplicity $\dim V_\CS$ and $\lambda_+$ in $V_\CM$ with multiplicity 
$\dim V_\CM/2$. 
We thus obtain 
\begin{eqnarray}
  && p_{\max}(d) \nonumber \\
  && = \eta_2 + \frac{1}{d_1d_2} 
          \left( 
              (\eta_1 - \eta_2) \dim V_\CS + \lambda_+\frac{\dim V_\CM }{2} 
          \right) \nonumber \\
  && = \frac{1}{2} + \frac{d+2}{6d} (\eta_1-\eta_2) + \frac{d-1}{3d}\sqrt{1-\eta_1 \eta_2}.
\end{eqnarray}

If $\eta_1 \le \eta_2$, the only positive eigenvalue of $\Delta$ is $\lambda_+$ in $V_\CM$;  
hence we obtain  
\begin{eqnarray}
  && p_{\max}(d)  
        = \eta_2 + \frac{1}{d_1d_2}\lambda_+\frac{\dim V_\CM }{2} \nonumber \\
  && = \frac{1}{2} - \frac{d+2}{6d} (\eta_1-\eta_2) + \frac{d-1}{3d}\sqrt{1-\eta_1 \eta_2}.
\end{eqnarray}

These two cases can be combined to yield a symmetric form of 
the maximum success identification probability for the 
general magnitude relation between $\eta_1$ and $\eta_2$: 
\begin{eqnarray}
  p_{\max} (d) = \frac{1}{2} + \frac{d+2}{6d} |\eta_1-\eta_2| + \frac{d-1}{3d}\sqrt{1-\eta_1 \eta_2}.
              \label{pmax}
\end{eqnarray}
The maximum is attained when the POVM element $E_1$ is given by $P_+$, the projector 
onto the subspace of positive eigenvalues of $\Delta$.  
The $p_{\max} (d)$ given by Eq.(\ref{pmax}) reproduces the result for the case $d=2$ obtained 
in Ref.\cite{Bergou06} and the one for arbitrary $d$ in Ref.\cite{Hayashi05}
when $\eta_1=\eta_2=1/2$. 

%%%%%%%%%%%%%%%%%%%%%%%%%%%%%%%%%%%%%%%%%%%%%%%%%%%%%%%%%%%%%%%%%%%%%%%%%%%%%
%%%%%%%%%%%%%%%%%%%%%%%%%%%%%%%%%%%%%%%%%%%%%%%%%%%%% Minimum-error identification by LOCC
%%%%%%%%%%%%%%%%%%%%%%%%%%%%%%%%%%%%%%%%%%%%%%%%%%%%%%%%%%%%%%%%%%%%%%%%%%%%%%%
\section{Minimum-error identification by LOCC}
Let us assume that each of the three systems 0, 1, and 2 , where the input state and the 
two reference states are prepared, consists of two subsystems. 
The state space of each system is represented by a tensor product 
$\BC^d=\BC^{d_a}\otimes \BC^{d_b}$, which is shared by Alice and Bob.  
Their task is to identify a given input bipartite state 
with one of the two bipartite reference states by means of local operations and classical 
communication. As in the preceding section, the two reference states are chosen 
independently and randomly from the pure state space $\BC^d$ in a unitary invariant way. 
Therefore, those bipartite states are generally entangled. 
The question is whether Alice and Bob can achieve the maximum mean identification success 
probability given by the global measurement scheme. In this section, we will show that 
this is possible by explicitly constructing a LOCC protocol which achieves it.  

The mean success probability is given by Eq.(\ref{pd}) in the preceding section. 
The optimal global POVM element $E_1$ is $P_+$, the projector onto 
the subspace of positive eigenvalues of $\Delta$ defined by Eq.(\ref{delta}).  
The projector $P_+$ does not apparently satisfy the conditions of LOCC, 
since the operator $\Delta$ is not of a separable form. 
However, it should be noticed that $\tr{E_1\Delta}$ remains the same if the support 
of $E_1$ contains states with zero eigenvalue of $\Delta$. It is this freedom that 
we will exploit in order to construct a POVM element $E_1$ that satisfies the 
LOCC conditions.  

We begin by rewriting the operator $\Delta$ of Eq.(\ref{deltaDA}) in terms of 
local operators of Alice and Bob. 
Note that the exchange operator $T(01)$, for example, can be written as 
$T(01)=T^{(a)}(01) \otimes T^{(b)}(01)$, where $T^{(a)}(01)$ is the operator that exchanges Alice's 
part of systems 0 and 1 and $T^{(b)}(01)$ is defined for Bob's part in the same way. Hereafter, 
we use the superscript $(a)$ or $(b)$ for an operator to indicate which space (of Alice or Bob)  
the operator acts on. Since we have 
\begin{eqnarray}
   D &=& D^{(a)}\otimes A^{(b)} + A^{(a)}\otimes D^{(b)},   \\
   A &=& D^{(a)}\otimes D^{(b)} + A^{(a)}\otimes A^{(b)},  
\end{eqnarray}
the operator $\Delta$ is expressed as
\begin{eqnarray}
   \Delta &=& \frac{1}{2} \Big(
               \eta_1-\eta_2 + D^{(a)}A^{(b)} + A^{(a)}D^{(b)}   \nonumber \\ 
          & & \ \ \ \ \ \ +(\eta_1-\eta_2) (D^{(a)}D^{(b)} + A^{(a)}A^{(b)}) \Big).
\end{eqnarray}

The task for Alice and Bob is to maximize $\tr{E_1^{{\rm L}}\Delta}$ with a POVM element  
$E_1^{{\rm L }}$ which satisfies LOCC conditions. 
We first construct a separable POVM $E_1^{{\rm L}}$ which 
attains the maximum value $\tr{P_+\Delta}$. This separable POVM $E_1^{{\rm L}}$ 
will then be shown to satisfy the LOCC conditions. 
Without loss of generality, we assume $\eta_1 \le \eta_2$ throughout this section, since the problem 
is symmetric with respect to $\rho_1$ and $\rho_2$. 

Suppose that Alice and Bob first determine the permutation symmetry of their systems by 
a projective measurement with projection operators 
$\{\CS_3^{(a)}, \CA_3^{(a)}, \CM_3^{(a)}\}$ and $\{\CS_3^{(b)}, \CA_3^{(b)}, \CM_3^{(b)}\}$, 
respectively.  
If one of them found that his or her system is totally symmetric or antisymmetric, 
it is easy for the other party to find the best strategy. 
For example, assume that Alice found her system to be totally symmetric.  
Knowing Alice's outcome, Bob performs a POVM measurement, which we denote by $x^{(b)}$. 
The contribution to $\tr{E_1^{{\rm L}}\Delta}$ is then given by
\begin{eqnarray*}
  \tr{\CS_3^{(a)}\otimes x^{(b)} \Delta} = \dim (V_\CS^{(a)}) \trm_b[x^{(b)}\Delta^{(b)}], 
\end{eqnarray*}
since $\CS_3^{(a)}D^{(a)}=0$ and $\CS_3^{(a)}A^{(a)}=\CS_3^{(a)}$. 
It is clear that the best strategy for Bob is to take the projector $P_+^{(b)}$ 
onto the positive-eigenvalue space of $\Delta^{(b)}$.  
Note that the positive-eigenvalue space of $\Delta^{(b)}$ is a subspace of 
$V_\CM^{(b)}$, since the eigenvalue of $\Delta^{(b)}$ in $V_\CS^{(b)}$ 
is $\eta_1 - \eta_2 (\le 0)$.  
In this case the contribution to $\tr{E_1^{{\rm L}} \Delta}$ is given by 
\begin{eqnarray*}
   \tr{\CS_3^{(a)} \otimes P_+^{(b)} \Delta} = 
           \frac{\lambda_+}{2} \dim V_{\CS}^{(a)} \dim V_{\CM}^{(b)}.
\end{eqnarray*}
If Alice's part is totally antisymmetric, the operator 
for Bob is given by 
\begin{eqnarray*}
  && \trm_a[\CA_3^{(a)}\Delta] =  \dim (V_\CA^{(a)}) \Delta^{'(b)}, \nonumber \\ 
  && \Delta^{'(b)} \equiv \frac{1}{2}\left( \eta_1-\eta_2 - D^{(b)} -(\eta_1-\eta_2)A^{(b)} \right).
\end{eqnarray*}
The operator $\Delta^{'(b)}$ differs from $\Delta^{(b)}$ only in the signs 
in front of $D^{(b)}$ and $A^{(b)}$. Its eigenvalues are 0 in $V_\CS^{(b)}$ 
and $\eta_1-\eta_2 (\le 0)$ in $V_\CA^{(b)}$. In $V_\CM^{(b)}$, the operator 
$\Delta^{'(b)}$ has eigenvalue $\lambda_-$ 
in the positive-eigenvalue subspace of $\Delta^{(b)}$ and $\lambda_+$ in the negative-eigenvalue 
subspace of $\Delta^{(b)}$. This implies that Bob's best POVM element is $P_-^{(b)}$, the projector 
onto the $\Delta^{(b)}$'s negative-eigenvalue subspace in $V_\CM^{(b)}$.
The contribution to $\tr{E_1^{{\rm L}} \Delta}$ in this case is given by
\begin{eqnarray*}
  \tr{\CA_3^{(a)} \otimes P_-^{(b)} \Delta} = 
           \frac{\lambda_+}{2} \dim V_{\CA}^{(a)} \dim V_{\CM}^{(b)}.
\end{eqnarray*}
The same argument also holds when Bob's system is totally symmetric or antisymmetric.  
Therefore, when the total state does not belong to $V_\CM^{(a)}\otimes V_\CM^{(b)}$, 
the whole contribution to $\tr{E_1^{{\rm L}} \Delta}$ is given by  
\begin{eqnarray}
  && \tr{ \left( \CS_3^{(a)}P_+^{(b)} + \CA_3^{(a)}P_-^{(b)}
                   +P_+^{(a)}\CS_3^{(b)} + P_-^{(a)}\CA_3^{(b)} \right) \Delta }
               \nonumber \\
  &=& \frac{1}{2}\lambda_+
              \Big( \dim V_\CS^{(a)}\dim V_\CM^{(b)}+\dim V_\CA^{(a)}\dim V_\CM^{(b)} 
                                    \nonumber \\
  && \hspace{5ex} +\dim V_\CM^{(a)}\dim V_\CS^{(b)}+\dim V_\CM^{(a)}\dim V_\CA^{(b)} \Big).
            \label{not_mm_part}
\end{eqnarray}

When the total state belongs to $V_\CM^{(a)}\otimes V_\CM^{(b)}$, construction of the best 
strategy for Alice and Bob is rather involved.  
First we introduce the following operators $X_1$ and $X_2$ for each of Alice's and Bob's spaces:  
\begin{eqnarray}
  X_1^{(\kappa)} &=& \frac{2}{\sqrt{3}} D^{(\kappa)},  \nonumber \\
  X_2^{(\kappa)}&=& 2A^{(\kappa)},\ \ \ (\kappa=a,b).
\end{eqnarray} 
Note that $X_1^{(\kappa)}$ and $X_2^{(\kappa)}$ anticommute and 
$(X_1^{(\kappa)})^2=(X_2^{(\kappa)})^2=1$ in the mixed symmetric space $V_\CM^{(\kappa)}$.  
The operator $\Delta$ in terms of $X_i^{(\kappa)}$ is not diagonal with respect to 
the index $i$. We further define rotated $X_i$'s in order to diagonalize $\Delta$ with 
respect to the index $i$: 
\begin{eqnarray}
  Y_1^{(\kappa)} &=& \cos\theta X_1^{(\kappa)} + \sin\theta X_2^{(\kappa)}, 
                          \nonumber \\
  Y_2^{(\kappa)} &=& -\sin\theta X_1^{(\kappa)} + \cos\theta X_2^{(\kappa)},\ \ (\kappa=a,b).
\end{eqnarray}
We find that $\Delta$ takes the following "diagonal" form: 
\begin{eqnarray}
  \Delta = \frac{1}{2} \left( \eta_1-\eta_2 + 
              \lambda_+Y_1^{(a)}Y_1^{(b)} + \lambda_-Y_2^{(a)}Y_2^{(b)} \right),
\end{eqnarray}
if we take
\begin{eqnarray*}
  \cos 2\theta &=& \frac{\eta_1-\eta_2}{2\sqrt{1-\eta_1\eta_2}}, \\
  \sin 2\theta &=& \frac{\sqrt{3}}{2\sqrt{1-\eta_1\eta_2}}. 
\end{eqnarray*}
The eigenvalues of $Y_i^{(\kappa)}$ are 1 and -1 with multiplicity $\dim V_\CM^{(\kappa)}/2$ 
since we have  
\begin{eqnarray}
  && (Y_1^{(\kappa)})^2 = 1,\ (Y_2^{(\kappa)})^2 = 1, \nonumber \\
  &&  Y_1^{(\kappa)}Y_2^{(\kappa)}+Y_2^{(\kappa)}Y_1^{(\kappa)} = 0.
\end{eqnarray}
The positive- and negative-eigenvalue subspaces of $Y_1^{(\kappa)}$ are transformed into 
each other by the operation of $Y_2^{(\kappa)}$ and vice versa. We should also notice that 
$|\lambda_-| \ge |\lambda_+|$ when $\eta_1 \le \eta_2$. 
These considerations imply that the optimal separate POVM element is given by 
$Q_+^{(a)}\otimes Q_-^{(b)} + Q_-^{(a)}\otimes Q_+^{(b)}$, where 
$Q_\pm^{(\kappa)}$ is the projector onto the positive- and negative-eigenvalue 
subspaces of $Y_2^{(\kappa)}$. The contribution to $\tr{E_1^{{\rm L}}\Delta}$ 
is found to be 
\begin{eqnarray}
  & & \tr{\left(Q_+^{(a)}\otimes Q_-^{(b)} + Q_-^{(a)}\otimes Q_+^{(b)}\right)\Delta}
                          \nonumber \\
  &=& \frac{1}{4}(\eta_1-\eta_2-\lambda_-) \dim V_\CM^{(a)} \dim V_\CM^{(b)} 
                          \nonumber \\
  &=& \frac{1}{4} \lambda_+ \dim V_\CM^{(a)} \dim V_\CM^{(b)},
               \label{mm_part}
\end{eqnarray}
where we used $\tr{Q_\pm^{(\kappa)}Y_1^{(\kappa)}}=0$. 

Thus the whole POVM element is given by 
\begin{eqnarray}
  E_1^{{\rm L}} &=& \CS_3^{(a)}P_+^{(b)} + \CA_3^{(a)}P_-^{(b)}
                   +P_+^{(a)}\CS_3^{(b)} + P_-^{(a)}\CA_3^{(b)} 
                                 \nonumber \\
                & &+Q_+^{(a)}Q_-^{(b)} + Q_-^{(a)}Q_+^{(b)}.   \label{E1L}
\end{eqnarray}
Adding Eqs.(\ref{not_mm_part}) and (\ref{mm_part}), we find that 
$\tr{E_1^{{\rm L}}\Delta}$ indeed attains the maximum value given by 
the global POVM element $E_1=P_+$:
\begin{eqnarray}
   \tr{E_1^{{\rm L}}\Delta} = \frac{1}{2}\lambda_+\dim{V_\CM} = \tr{P_+\Delta}.
\end{eqnarray} 
To show the above equality, we used the relation
\begin{eqnarray}
  \dim{V_\CM} &=& \dim{V_\CS^{(a)}}\dim{V_\CM^{(b)}}+\dim{V_\CM^{(a)}}\dim{V_\CS^{(b)}}
                                   \nonumber \\
              & &+\dim{V_\CA^{(a)}}\dim{V_\CM^{(b)}}+\dim{V_\CM^{(a)}}\dim{V_\CA^{(b)}}
                                   \nonumber \\
              & &+\frac{1}{2}\dim{V_\CM^{(a)}}\dim{V_\CM^{(b)}},
                              \label{dim_relation}
\end{eqnarray}
which can be readily verified by a straightforward calculation. 
This relation can be also understood from the viewpoint of inner (Kronecker) products 
of two representations of the symmetric group of order 3: the product of two mixed 
symmetric representations contains the totally symmetric and antisymmetric 
representations in addition to the mixed symmetric representation, whereas the product 
of the totally (anti)symmetric representation and the mixed symmetric representation is 
the mixed symmetric representation.  

We can show that the POVM element $E_1^{{\rm L}}$ given in Eq.(\ref{E1L}) 
can be implemented with an LOCC protocol. 
First Alice and Bob determine which permutation symmetries each one's local state has; 
totally symmetric, totally antisymmetric, or mixed symmetric. 
If one of them finds that his or her state is totally symmetric or antisymmetric and 
the other party's state is mixed symmetric, this party with the mixed symmetric state 
performs the measurement by the projectors $\{P_+^{(\kappa)},P_-^{(\kappa)}\}$. 
They conclude that the input state is $\rho_1$ if the combination of their 
outcomes is either (symmetric, $P_+$) or (antisymmetric, $P_-$). Otherwise 
they conclude that the input state is $\rho_2$. When Alice and Bob find 
both the local states are mixed symmetric, they perform the projection 
measurement by $\{ Q_+^{(\kappa)},Q_-^{(\kappa)}\}$. 
They conclude that the input state is $\rho_1$, only when the combination of outcomes is 
"($+$,$-$)". 

Thus we conclude that minimum-error pure-state identification with any prior 
occurrence probability can be optimally performed within the LOCC scheme.

%%%%%%%%%%%%%%%%%%%%%%%%%%%%%%%%%%%%%%%%%%%%%%%%%%%%%%%%%%
%%%%%%%%%%%%%%%%%%%%%%%%%%%%%%%%%%%%%%%  Unambiguous identification and symmetries of POVM
%%%%%%%%%%%%%%%%%%%%%%%%%%%%%%%%%%%%%%%%%%%%%%%%%%%%%%%%%%
\section{Unambiguous identification and symmetries of POVM}
In this section, we precisely formulate the unambiguous identification problem of two pure 
states, and rederive the optimal success probability attainable by 
the global measurement scheme, though the result has been reported in Refs.\cite{Bergou05,Hayashi06}. 
The purpose of this section is to explain two important symmetries 
of the measurement scheme of this problem, which will also play a crucial role in 
determining the optimal probability by the LOCC scheme in the next section. In this and the next 
sections, we assume equal prior probabilities for the two reference states.  

\subsection{Problem}
In the unambiguous identification problem, the task is to unambiguously identify the input pure state 
$\rho$ with one of the two pure reference states $\rho_1$ and $\rho_2$. 
Our measurement can produce three outcomes $\mu=1,2,0$. If the outcome is $\mu(=1,2)$, we are certain 
that the input state $\rho$ is $\rho_\mu$, and outcome 0 means that we are not certain 
about the identity of the input; this is called an inconclusive result. 
Let us introduce a POVM, $\{E_\mu \}_{\mu=0,1,2}$ corresponding to the three measurement outcomes.   
The mean success probability of identification is then given by
\begin{eqnarray}
  p &=& \frac{1}{2} \Big< \tr{E_1\rho_1(0)\rho_1(1)\rho_2(2)}
                  \nonumber \\
    & & \hspace{8ex} +\tr{E_2\rho_2(0)\rho_1(1)\rho_2(2)} \Big>.
                   \label{p1}
\end{eqnarray}
The condition that we are not allowed to make a mistake imposes the following no-error conditions 
on $E_1$ and $E_2$: 
\begin{eqnarray}
   \left\{ 
   \begin{array}{c}
      \tr{E_1\rho_2(0)\rho_1(1)\rho_2(2)} = 0, \\
      \tr{E_2\rho_1(0)\rho_1(1)\rho_2(2)} = 0, \\ 
   \end{array}
   \right. 
   \ \ \mbox{for any $\rho_1$ and $\rho_2$}.
                          \label{no-error1}
\end{eqnarray}

In what follows, we will optimize the mean success probability of Eq.(\ref{p1}) under the 
no-error conditions Eq.(\ref{no-error1}). 
Performing the average over the reference states by Eq.(\ref{formula}), 
we write the mean success probability (\ref{p1}) as 
\begin{eqnarray}
  p(d) = \frac{1}{2d_2d_1}\left( \tr{E_1\CS(01)} + \tr{E_2\CS(02)} \right), 
           \label{pS}
\end{eqnarray}
where $\CS(01)$ and $\CS(02)$ are the projectors onto the totally symmetric subspaces of 
spaces $0 \otimes 1$ and $0 \otimes 2$, respectively.  
Averaging the no-error conditions of Eq.(\ref{no-error1}), we obtain
\begin{eqnarray}
  \tr{E_1\CS(02)} = 0,\ \ \tr{E_2\CS(01)}=0.    \label{no-error2}
\end{eqnarray}
Since $E_1$ and $\CS(02)$ are both positive operators, the above conditions imply that  
the supports of them are orthogonal to each other. The same is true for 
$E_2$ and $\CS(01)$. The no-error conditions are thus equivalent to 
\begin{eqnarray}
  && E_1 \CS(02) = \CS(02) E_1 = 0, \label{no-error31} \\
  && E_2 \CS(01) = \CS(01) E_2 = 0. \label{no-error32}
\end{eqnarray}

\subsection{Symmetries of POVM}
The set of POVMs satisfying the no-error conditions is convex; 
if two POVMs $E_\mu$ and $E'_\mu$ respect the no-error conditions, 
so does their convex linear combination $rE_\mu+(1-r)E'_\mu$ for any 
$ 0 \le r \le 1$.
The resulting success probability is also a convex combination, 
$p(rE+(1-r)E') = rp(E) + (1-r)p(E')$, with an obviously abbreviated notation. 
It is this convexity of the POVM that we exploit in order to impose two symmetries 
on the optimal POVM without loss of generality. 

First we consider the exchange symmetry between systems 1 and 2.
For an optimal POVM $F_\mu$, we define another POVM by 
\begin{eqnarray}
  && F'_1 = T(12) F_2 T(12),   \nonumber \\
  && F'_2 = T(12) F_1 T(12),   \nonumber \\
  && F'_0 = T(12) F_0 T(12),   \nonumber 
\end{eqnarray}
where $T(12)$ is the exchange operator between systems 1 and 2.
The POVM $F'_\mu$ is clearly legitimate and optimal. Furthermore,
a new POVM $ E_\mu = \frac{1}{2} \left( F_\mu + F'_\mu \right) $, 
which is a convex linear combination of $F_\mu$ and $F'_\mu$, 
is also optimal and satisfies the exchange symmetry between systems 1 and 2:
\begin{eqnarray}
   && E_1=T(12)E_2T(12),   \nonumber \\
   && E_2=T(12)E_1T(12),   \nonumber \\
   && E_0=T(12)E_0T(12).   \label{exchange_symmetry}
\end{eqnarray}

The second important symmetry is the unitary symmetry of the 
distribution of the reference states. If a POVM $F_\mu$ is optimal, 
another POVM defined by 
\begin{eqnarray*}
   U^{\otimes 3}F_\mu (U^{\otimes 3})^{-1}, 
\end{eqnarray*} 
is also legitimate and optimal for arbitrary unitary operator $U$.
We now construct a POVM by 
\begin{eqnarray}
    E_\mu = \int dU U^{\otimes (3)} F_\mu (U^{\otimes (3)})^{-1}, 
\end{eqnarray}
where $dU$ is the normalized positive invariant measure of the group $U(d)$.
The POVM $E_\mu$ is clearly a legitimate and optimal POVM.  
We can show that $E_\mu$ commutes with $U^{\otimes 3}$ for any unitary $U$:
\begin{eqnarray}
   U^{\otimes 3} E_\mu &=&  \int dU' (UU')^{\otimes (3)} F_\mu (UU'^{\otimes (3)})^{-1}U^{\otimes 3}
                         \nonumber \\
     &=& \int dU' U'^{\otimes (3)} F_\mu (U'^{\otimes (3)})^{-1} U^{\otimes 3}
                         \nonumber \\
     &=& E_\mu U^{\otimes 3},
\end{eqnarray}
which means that $E_\mu$ is a scalar with respect to the group $U(d)$. 
Thus we can assume that the optimal POVM satisfies the exchange symmetry of 
Eq.(\ref{exchange_symmetry}) and is scalar with respect to the group $U(d)$.

By the exchange symmetry, the mean success probability to be optimized 
takes the form 
\begin{eqnarray}
 p = \frac{1}{d_2d_1}\tr{E_1\CS(01)}, 
                                       \label{pS_symmetric}
\end{eqnarray}
where $E_1$ is a unitary scalar and subject to the conditions 
\begin{eqnarray}
  E_1 \ge 0,\ 1 \ge E_1+T(12)E_1T(12), 
                 \label{positivity_symmetric}
\end{eqnarray}
in addition to the no-error conditions given by Eq.(\ref{no-error31}).

\subsection{Optimal identification probability}
As in the two preceding sections, we decompose the total space into three subspaces 
$V_\CS$, $V_\CA$, and $V_\CM$, 
according to the symmetry with respect to system permutations. 
Under the unitary transformation $U^{\otimes 3}$, the three subspaces $V_\CS$, $V_\CA$, and $V_\CM$ 
are clearly invariant since $U^{\otimes 3}$ commutes with system permutations.  
Furthermore, it is known that $U^{\otimes 3}$ acts on $V_\CS$ and $V_\CA$ irreducibly, 
and the mixed symmetric space $V_\CM$ contains two $(\dim V_\CM /2)$-dimensional irreducible 
representations of the group $U(d)$ \cite{Hamermesh62}. 
Now, suppose that a positive operator $E$ is a unitary scalar; $E$ commutes with 
$U^{\otimes 3}$. If $\tr{E\CS(02)}=0$, Schur's lemma requires that $E$ be a linear combination of 
two projection operators $\CA_3$ and $\CM_3 \CA(02)$, with $\CA(02)=1-\CS(02)$ being 
the projector onto the antisymmetric subspace of $0 \otimes 2$. 
Similarly, if $\tr{E\CA(02)}=0$, $E$ is a linear combination of $\CS_3$ and $\CM_3 \CS(02)$. 
These facts will be used also in the next section.

With these considerations, we can determine the operator form of the POVM element $E_1$.  
The operator $E_1$ is given by a linear combination of $\CA_3$ and $\CM_3 \CA(02)$ since 
$E_1$ is a unitary scalar and satisfies the no-error condition $\tr{E_1\CS(02)}=0$. 
But $\CA_3$ does not contribute to the mean success probability of Eq.(\ref{pS_symmetric}). 
Thus, without loss of generality, we can write 
\begin{eqnarray}
  E_1 = \alpha \CM_3 \CA(02),
\end{eqnarray}
where $\alpha$ is a positive coefficient. The range of $\alpha$ is restricted by the positivity 
of $E_0$, which is written as 
\begin{eqnarray*}
    E_1 + T(12)E_1T(12) &=& \alpha\CM_3(\CA(02)+\CA(01))   \nonumber \\
                        &=& \alpha\CM_3\left( 1-A \right) \le 1.
\end{eqnarray*}
This requires $\alpha \le 2/3 $, since the operator $A$ defined in Eq.(\ref{A}) has 
eigenvalues $\pm 1/2$ in $V_\CM$. 
Clearly, the mean success probability attains its maximum when $\alpha$ takes the 
largest possible value, $2/3$. The optimal POVM is thus given by 
\begin{eqnarray}
  E_1 &=& \frac{2}{3}\CM_3\CA(02),   \nonumber \\
  E_2 &=& \frac{2}{3}\CM_3\CA(01),  \nonumber \\
  E_0 &=& \frac{1}{3}\CM_3(1+2A)+\CS_3+\CA_3. 
\end{eqnarray}
In order to obtain the optimal probability, we need trace $\tr{\CM_3\CA(02)\CS(01)}$, 
which is calculated as follows: 
\begin{eqnarray}
 &&  \tr{\CM_3\CA(02)\CS(01)}     \nonumber \\
 &=& \frac{1}{4}\tr{\CM_3 (1+T(01)-T(02)-T(02)T(01))}  \nonumber \\
 &=& \frac{1}{2}\tr{\CM_3 D^2} = \frac{3}{8}\dim V_\CM, 
\end{eqnarray}
where we used $\tr{\CM_3T(01)}=\tr{\CM_3T(02)}=0$.
Using the explicit expressions for the dimensions, we finally obtain 
the optimal mean success probability of unambiguous identification:
\begin{eqnarray}
  p_{\max}= \frac{d-1}{3d}. \label{global_p}
\end{eqnarray}

%%%%%%%%%%%%%%%%%%%%%%%%%%%%%%%%%%%%%%%%%%%%%%%%%%%%%%%%%%
%%%%%%%%%%%%%%%%%%%%%%%%%%%%%%%%%%%%%%%  Local unambiguous identification
%%%%%%%%%%%%%%%%%%%%%%%%%%%%%%%%%%%%%%%%%%%%%%%%%%%%%%%%%%
\section{Local unambiguous identification}
Let us now assume that each of the three systems consists of two subsystems shared by 
Alice and Bob and its state space is represented by a tensor product 
$\BC^d=\BC^{d_a} \otimes \BC^{d_b}$.  
The task of Alice and Bob is to unambiguously identify a given input state 
by means of LOCC with one of the two reference states. 

\subsection{Separable POVM and symmetries}
Any POVM $E_\mu^{{\rm L}}$ which satisfies the LOCC conditions has a separable 
form:  
\begin{eqnarray}
    E_\mu^{{\rm L}} = \sum_i E_{\mu i}^{(a)} \otimes E_{\mu i}^{(b)},\ \ \ (\mu=0,1,2), 
\end{eqnarray}
where
\begin{eqnarray}
     E_{\mu i}^{(a)} \ge 0, \ \ E_{\mu i}^{(b)} \ge 0,\ \ \sum_\mu E_\mu^{{\rm L}} = 1.
\end{eqnarray}
It is known that there exist separable POVMs that do not satisfy the LOCC conditions \cite{Bennett99}. 
We will first optimize the success probability within the separable class of POVMs, and 
then show that the obtained optimal separable POVM can be implemented by a LOCC protocol.
 
Note that a convex linear combination of separable POVMs is again separable, and the no-error 
conditions $\tr{E_1^{{\rm L}}\CS(02)}=\tr{E_2^{{\rm L}}\CS(01)}=0$ are also preserved. 
This enables us to impose two symmetries on the optimal separable POVM as in the preceding 
section. We begin with the exchange symmetry for systems 1 and 2. Since 
$T(12)=T^{(a)}(12) \otimes T^{(b)}(12)$, it is clear that $T(12)E_\mu^{{\rm L}} T(12)$ is separable if 
$E_\mu^{{\rm L}}$ is separable:
\begin{eqnarray*}
  && T(12)E_\mu^{{\rm L}} T(12)   \nonumber \\ 
  &=& \sum_i T^{(a)}(12)E_{\mu i}^{(a)}T^{(a)}(12) \otimes 
                             T^{(b)}(12)E_{\mu i}^{(b)}T^{(b)}(12).
\end{eqnarray*}
Therefore, we can impose on separable POVMs the same exchange symmetry as the one given in 
Eq.(\ref{exchange_symmetry}):
\begin{eqnarray}
  E_1^{{\rm L}} &=& T(12) E_2^{{\rm L}} T(12), \nonumber \\
  E_2^{{\rm L}} &=& T(12) E_1^{{\rm L}} T(12), \nonumber \\
  E_0^{{\rm L}} &=& T(12) E_0^{{\rm L}} T(12). 
\end{eqnarray}

For the unitary symmetry, we notice that $U^{\otimes 3} F_\mu^{{\rm L}} (U^{\otimes 3})^{-1}$ is not 
generally separable for a separable POVM 
$F_\mu^{{\rm L}} = \sum_i F_{\mu i}^{(a)} \otimes F_{\mu i}^{(b)}$. 
However, this is true if $U$ is a tensor product of two unitaries as $U=u^{(a)} \otimes v^{(b)}$:
\begin{eqnarray*}
   & & U^{\otimes 3} F_\mu^{{\rm L}} (U^{\otimes 3})^{-1} \nonumber \\
   &=& \sum_i  u^{(a)\otimes 3} F_{\mu i}^{(a)} (u^{(a)\otimes 3})^{-1} \otimes
            v^{(b)\otimes 3} F_{\mu i}^{(b)} (v^{(b)\otimes 3})^{-1}.
\end{eqnarray*}
For the class of this separable $U$, we can repeat the argument given in the preceding section.
Assume that a separable POVM $F_\mu^{{\rm L}} $ is optimal.  
Integrating over $u^{(a)}$ and $v^{(b)}$ with the invariant measure, we obtain 
\begin{eqnarray*}
   E_\mu^{{\rm L}} &\equiv& \int du^{(a)}dv^{(b)}\left(u^{(a) \otimes 3} v^{(b) \otimes 3} \right)
                       F_\mu^{{\rm L}}    \left(u^{(a) \otimes 3} v^{(b) \otimes 3} \right)^{-1}  
                                    \nonumber \\
         &=& \sum_i E_{\mu i}^{(a)}\otimes E_{\mu i}^{(b)}, 
\end{eqnarray*}
where $E_{\mu i}^{(a)}$ and $E_{\mu i}^{(b)}$ are given by 
\begin{eqnarray*}
   E_{\mu i}^{(a)} &=& \int du^{(a)} u^{(a)\otimes 3} F_{\mu i}^{(a)} (u^{(a)\otimes 3})^{-1}, 
                        \nonumber \\
   E_{\mu i}^{(b)} &=& \int dv^{(b)} v^{(b)\otimes 3} F_{\mu i}^{(b)} (v^{(b)\otimes 3})^{-1}. 
\end{eqnarray*} 
The POVM $ E_\mu^{{\rm L}} $ obtained in this way is again separable and optimal. 
Furthermore, it is easy to see that $E_{\mu i}^{(a)}$ and $E_{\mu i}^{(b)}$ are both 
unitary scalar: for any unitaries $u^{(a)}$ and $v^{(b)}$, we have 
\begin{eqnarray}
    [E_{\mu i}^{(a)}, u^{(a)}] = 0,\ \ \ [E_{\mu i}^{(b)}, v^{(b)}]=0.
\end{eqnarray}

Let us closely examine the no-error conditions for separable POVMs. 
The global projector $\CS(02)$ is decomposed by local symmetry projectors as 
follows:                        
\begin{eqnarray*}
  \CS(02) = \CS^{(a)}(02) \otimes \CS^{(b)}(02)+\CA^{(a)}(02) \otimes \CA^{(b)}(02).
\end{eqnarray*}
The no-error condition $\tr{E_1^{{\rm L}}\CS(02)}=0$ is then expressed as  
\begin{eqnarray*}
  && \sum_i \Big( \tr{E_{1 i}^{(a)}\CS^{(a)}(02)}\tr{E_{1 i}^{(b)}\CS^{(b)}(02)} 
                           \nonumber \\
  && \hspace{2ex} +\tr{E_{1 i}^{(a)}\CA^{(a)}(02)}\tr{E_{1 i}^{(b)}\CA^{(b)}(02)} 
            \Big) = 0.
\end{eqnarray*}                     
In this equation, all terms are non-negative, implying that each term should vanish. 
Therefore, for each $i$, we have two possibilities: one is 
\begin{eqnarray*}
   \tr{E_{1 i}^{(a)}\CS^{(a)}(02)}=0,\ \mbox{and}\ \tr{E_{1 i}^{(b)}\CA^{(b)}(02)} =0,
                       \label{one_possibility}
\end{eqnarray*}
and the other is 
\begin{eqnarray*}
   \tr{E_{1 i}^{(a)}\CA^{(a)}(02)}=0,\ \mbox{and}\ \tr{E_{1 i}^{(b)}\CS^{(b)}(02)} =0.
                       \label{other_possibility}
\end{eqnarray*}
Note that other combinations like 
\begin{eqnarray*}
   \tr{E_{1 i}^{(a)}\CS^{(a)}(02)}=0,\ \mbox{and}\ \tr{E_{1 i}^{(a)}\CA^{(a)}(02)} =0,
\end{eqnarray*}
do not occur as this would imply that $E_{1 i}^{(a)}$ or $E_{1 i}^{(b)}$ is identically zero. 
From the no-error condition for $E_2^{{\rm L}}$, we obtain the similar conditions 
for its components $E_{2 i}^{(a)}$ and $E_{2 i}^{(b)}$.

\subsection{Possible operator form of separable POVM}

As in the preceding section, we can show that a positive operator $E^{(p)}$ on space $V^{(p)}(p=a,b)$
which is a unitary scalar and satisfies $\tr{E^{(p)}\CS^{(p)}(02)}=0$ 
is a linear combination of $\CA_3^{(p)}$ and $\CM_3^{(p)}\CA^{(p)}(02)$. Similarly, 
if $E^{(p)}$ satisfies $\tr{E^{(p)}\CA^{(p)}(02)}=0$, then 
$E^{(p)}$ can be written as a linear combination of $\CS_3^{(p)}$ and $\CM_3^{(p)}\CS^{(p)}(02)$.
Now we can write the possible form of the separable $E_1^{{\rm L}}$ which has the unitary symmetry and 
satisfies the no-error conditions: 
\begin{eqnarray}
   E_1^{{\rm L}} &=& 
       \alpha_1 \CS_3^{(a)} \otimes \CM_3^{(b)}\CA^{(b)}(02)
      +\alpha_2 \CA_3^{(a)} \otimes \CM_3^{(b)}\CS^{(b)}(02)   
                               \nonumber \\
   &+& \alpha_3 \CM_3^{(a)}\CS^{(a)}(02) \otimes \CA_3^{(b)}
      +\alpha_4 \CM_3^{(a)}\CA^{(a)}(02) \otimes \CS_3^{(b)}  
                               \nonumber \\
   &+& \beta_1 \CM_3^{(a)}\CS^{(a)}(02) \otimes \CM_3^{(b)}\CA^{(b)}(02)    
                               \nonumber \\              
   &+& \beta_2 \CM_3^{(a)}\CA^{(a)}(02) \otimes \CM_3^{(b)}\CS^{(b)}(02),  
                        \label{possible_E1}
\end{eqnarray}
where $\alpha_1,\alpha_2,\alpha_3,\alpha_4$ and $\beta_1,\beta_2$ are non-negative 
coefficients. The operators $\CS_3 \otimes \CA_3$ and $\CA_3 \otimes \CS_3$ are not 
included in $E_1^{{\rm L}}$, since the corresponding outcomes do not arise; the total state  
contains no totally antisymmetric component.  
$E_2^{{\rm L}}$ and $E_0^{{\rm L}}$ are given by 
$E_2^{{\rm L}} = T(12)E_1^{{\rm L}}T(12)$ and $E_0^{{\rm L}}=1-E_1^{{\rm L}}-E_2^{{\rm L}}$.

We have now found the possible operator form of a separable POVM that respects the no-error conditions. 
The remaining requirement on the POVM is the positivity of $E_0^{{\rm L}}$, which restricts 
the range of the coefficients $\alpha$ and $\beta$. The positivity of $E_0^{{\rm L}}$ 
is equivalent to $E_1^{{\rm L}}+E_2^{{\rm L}} \le 1$. We will separately check this inequality 
in each of all subspaces of the permutation symmetry. 

In the subspace $V_{\CS}^{(a)} \otimes V_{\CM}^{(b)}$, the relevant part of $E_1^{{\rm L}}+E_2^{{\rm L}}$ 
is written as 
\begin{eqnarray*}
  && \alpha_1 \left( \CS_3^{(a)} \otimes \CM_3^{(b)}\CA^{(b)}(02) +
                      \CS_3^{(a)} \otimes \CM_3^{(b)}\CA^{(b)}(01)  \right)
                                 \nonumber \\
  &&=\alpha_1 \CS_3^{(a)} \otimes \CM_3^{(b)} (1-A^{(b)}) .
\end{eqnarray*}
This part should be smaller than the projector onto the subspace, $\CS_3^{(a)} \otimes \CM_3^{(b)}$. 
Here, $A^{(b)}$ is defined to be $(T^{(b)}(01)+T^{(b)}(02))/2$ in the same way for the global operator $A$ defined by Eq.(\ref{A}). The eigenvalues of $A^{(b)}$ in $V_{\CM}^{(b)}$ are $1/2$ and $-1/2$. 
Therefore, we obtain $\alpha_1 \le 2/3$. Similarly, we obtain $\alpha_4 \le 2/3$ from the inequality 
in the subspace $V_{\CM}^{(a)} \otimes V_{\CS}^{(b)}$.

The inequality $E_1^{{\rm L}}+E_2^{{\rm L}} \le 1$ in $V_{\CA}^{(a)} \otimes V_{\CM}^{(b)}$ 
takes the form:  
\begin{eqnarray*}
  & & \alpha_2 \left( \CA_3^{(a)} \otimes \CM_3^{(b)}\CS^{(b)}(02) +
                      \CA_3^{(a)} \otimes \CM_3^{(b)}\CS^{(b)}(01)  \right)
                                 \nonumber \\
  & &=\alpha_2 \CA_3^{(a)} \otimes \CM_3^{(b)} (1+A^{(b)}) \le \CA_3^{(a)} \otimes \CM_3^{(b)},  
\end{eqnarray*} 
which requires that $\alpha_2 \le 2/3$.  In the same way, we obtain $\alpha_3 \le 2/3$  
from the inequality in $V_{\CM}^{(a)} \otimes V_{\CA}^{(b)}$. Thus all the four coefficients $\alpha$ 
should be less than or equal to 2/3. 

It is not straightforward to find allowed ranges of $\beta_1$ and $\beta_2$ from 
the inequality in $V_\CM^{(a)} \otimes V_\CM^{(b)}$.  
In the space $V_\CM^{(a)} \otimes V_\CM^{(b)}$, we define an operator $X$ to be 
\begin{eqnarray}
  X \equiv \beta_1 \left( \CS^{(a)}(02) \CA^{(b)}(02) 
                           +\CS^{(a)}(01) \CA^{(b)}(01) \right)\  &&
                                      \nonumber \\
   +\beta_2 \left( \CA^{(a)}(02) \CS^{(b)}(02) 
                            +\CA^{(a)}(01) \CS^{(b)}(01) \right), &&
\end{eqnarray}
which is the part of $E_1^{{\rm L}}+E_2^{{\rm L}}$ contributing to the space 
$V_\CM^{(a)} \otimes V_\CM^{(b)}$. 
We need to find the greatest eigenvalue of $X$, since the inequality implies 
$X \le \CM_3^{(a)} \otimes \CM_3^{(b)}$. 
It should be understood that we are working in subspace $V_\CM^{(a)} \otimes V_\CM^{(b)}$, and 
the projectors $\CM_3^{(a)}$ and $\CM_3^{(b)}$ will be omitted. 
In terms of $A^{(p)}$ and $D^{(p)}$, the operator $X$ is expressed as 
\begin{eqnarray*}
  X = \beta \left( 1 - A^{(a)} \otimes A^{(b)} - D^{(a)} \otimes D^{(b)} \right)\ &&
                      \nonumber \\
     +\delta \left( 1^{(a)} \otimes A^{(b)} - A^{(a)} \otimes 1^{(b)} \right), &&
\end{eqnarray*}
where $\beta=\frac{1}{2}(\beta_1+\beta_2)$ and $\delta=\frac{1}{2}(\beta_1-\beta_2)$. 
In order to diagonalize $X$, it is convenient to introduce the basis in which $A^{(p)}(p=a,b)$ 
is diagonal:
\begin{eqnarray*}
  A^{(p)}\ket{m+} = \frac{1}{2} \ket{m+},\ A^{(p)}\ket{m-} = -\frac{1}{2}\ket{m-},  &&  
                   \nonumber \\
    D^{(p)}\ket{m+} = \frac{\sqrt{3}}{2}\ket{m-},\ D^{(p)}\ket{m-} = \frac{\sqrt{3}}{2}\ket{m+},  &&  
\end{eqnarray*}  
where $m=1,2,\ldots,\dim (V_\CM^{(p)})/2$. The bipartite state 
$\ket{m\pm} \otimes \ket{m'\pm}$ for a given set of $m$ and $m'$ will be 
written as $\ket{\!\pm\pm}$ for simplicity.   

In this basis, two eigenvalues of $X$ are easily found by inspection:
\begin{eqnarray*}
  && X ( \ket{\!++}+\ket{\!--} ) = 0,  \\
  && X ( \ket{\!++}-\ket{\!--} ) = \frac{3}{2}\beta ( \ket{\!++}-\ket{\!--} ).
\end{eqnarray*}
States $\ket{\!+-}$ and $\ket{\!-+}$ are transformed by $X$ as 
\begin{eqnarray*}
  X \ket{\!+-} &=& \left(\frac{5}{4}\beta-\delta \right) \ket{\!+-} 
                         -\frac{3}{4}\beta\ket{\!-+} ,
                                           \nonumber \\ 
  X \ket{\!-+} &=& \left(\frac{5}{4}\beta+\delta \right) \ket{\!-+} 
                         -\frac{3}{4}\beta\ket{\!+-}.
\end{eqnarray*}
The other two eigenvalues are determined by diagonalizing the $2 \times 2$ matrix corresponding to 
the above transformation and found to be  
\begin{eqnarray*}
     \gamma_{\pm} = \frac{5}{4}\beta \pm \sqrt{ \frac{9}{16}\beta^2+\delta^2}.
\end{eqnarray*}
Of the four eigenvalues, the greatest one is $\gamma_+$. 
The positivity of $E_0^{{\rm L}}$ thus requires that the positive coefficients $\beta_1$ and 
$\beta_2$ should satisfy the condition:
\begin{eqnarray}
  \frac{5}{4}\beta + \sqrt{ \frac{9}{16}\beta^2+\delta^2} \le 1, 
\end{eqnarray}
where $\beta=\frac{1}{2}(\beta_1+\beta_2)$ and $\delta=\frac{1}{2}(\beta_1-\beta_2)$.

\subsection{Maximum success probability by separable POVM}
Now that we have the possible form of the separable POVM $E_\mu^{{\rm L}}$ and the conditions 
for the coefficients in it, we can optimize the mean success probability given by 
\begin{eqnarray}
  p^{{\rm L}} = \frac{1}{d_2d_1} \tr{E_1^{{\rm L}}\CS(01)}.
\end{eqnarray}
The trace $\tr{E_1^{{\rm L}}\CS(01)}$ can be calculated by decomposing the trace into traces  
in the subsystems as  
\begin{eqnarray*}
   & & \tr{E_1^{{\rm L}}\CS(01)}  \\
   &=&  \tr{E_1^{{\rm L}}  
         \left( \CS^{(a)}(01)  \CS^{(b)}(01) + \CA^{(a)}(01)  \CA^{(b)}(01) \right) }. 
\end{eqnarray*}
We must calculate many traces in subsystems, for which 
the following formulas can be used ($p=a,b$):
\begin{eqnarray*}
 & & \trm \CM_3^{(p)}\CS^{(p)}(02)\CS^{(p)}(01) = \trm \CM_3^{(p)}\CA^{(p)}(02)\CA^{(p)}(01)  \\
 & & = \frac{1}{2}\tr{\CM_3^{(p)}(A^{(p)})^2} = \frac{1}{8}\dim V_\CM^{(p)},   \\
 & & \trm \CM_3^{(p)}\CS^{(p)}(02)\CA^{(p)}(01) = \trm \CM_3^{(p)}\CA^{(p)}(02)\CS^{(p)}(01)  \\
 & & = \frac{1}{2}\tr{\CM_3^{(p)}(D^{(p)})^2} = \frac{3}{8}\dim V_\CM^{(p)}. 
\end{eqnarray*}
The result is given by 
\begin{eqnarray*}
 & & \tr{E_1^{{\rm L}}\CS(01)}      \\ 
 &=& \frac{3}{8} \Big( 
         \alpha_1 \dim V_\CS^{(a)} \dim V_\CM^{(b)} + \alpha_2 \dim V_\CA^{(a)} \dim V_\CM^{(b)} \\ 
 & &    +\alpha_3 \dim V_\CM^{(a)} \dim V_\CA^{(b)} + \alpha_4 \dim V_\CM^{(a)} \dim V_\CS^{(b)}
                 \Big)            \\
 & &   +\frac{3}{32}(\beta_1+\beta_2)\dim V_\CM^{(a)} \dim V_\CM^{(b)}.
\end{eqnarray*}
It is clear that we should take the largest possible value 2/3 for 
the coefficients $\alpha$ in order to maximize $\tr{E_1^{{\rm L}}\CS(01)}$.  
For $\beta_1$ and $\beta_2$, note that $\tr{E_1^{{\rm L}}\CS(01)}$ contains $\beta$'s in the 
form of $\beta_1+\beta_2$, and we can use the following inequalities:
\begin{eqnarray}
   \beta_1+\beta_2= 2\beta \le \frac{5}{4}\beta + \sqrt{ \frac{9}{16}\beta^2+\delta^2} \le 1.
\end{eqnarray}
Evidently, $\beta_1+\beta_2$ takes the maximum value 1 only when $\beta_1=\beta_2=1/2$. 
The maximum value of $\tr{E_1^{{\rm L}}\CS(01)}$ with separable POVM is thus given by 
\begin{eqnarray}
 & & \tr{E_1^{{\rm L}}\CS(01)}      \nonumber    \\ 
 &=& \frac{1}{4} \Big( 
         \dim V_\CS^{(a)} \dim V_\CM^{(b)} + \dim V_\CA^{(a)} \dim V_\CM^{(b)}  \nonumber \\ 
 & &    +\dim V_\CM^{(a)} \dim V_\CA^{(b)} + \dim V_\CM^{(a)} \dim V_\CS^{(b)}
                                    \nonumber    \\
 & &   +\frac{3}{8} \dim V_\CM^{(a)} \dim V_\CM^{(b)} \Big).   \label{separable_trace_ES}
\end{eqnarray}
On the other hand, we have 
\begin{eqnarray} 
  \tr{E_1\CS(01)}= \frac{1}{4}\dim V_\CM,
\end{eqnarray}
for the global POVM element $E_1$. Using the relation of dimensions given in 
Eq.(\ref{dim_relation}), we thus conclude that $\tr{E_1^{{\rm L}}\CS(01)} < \tr{E_1\CS(01)}$, 
implying that any separable POVM cannot attain the maximum unambiguous identification probability 
by the global measurement scheme.  
 
\subsection{LOCC protocol}
Thus, the optimal separable POVM element $E_1^{{\rm L}}$ is given by Eq.(\ref{possible_E1}) with 
$\alpha_i=2/3$ and $\beta_i=1/2$:
\begin{eqnarray}
  E_1^{{\rm L}} &=&  \frac{2}{3} \Big( 
           \CS_3^{(a)} \otimes \CM_3^{(b)}\CA^{(b)}(02)
          +\CA_3^{(a)} \otimes \CM_3^{(b)}\CS^{(b)}(02)   \nonumber \\
      & & +\CM_3^{(a)}\CS^{(a)}(02) \otimes \CA_3^{(b)}
          +\CM_3^{(a)}\CA^{(a)}(02) \otimes \CS_3^{(b)} \Big)
                               \nonumber \\
      & & +\frac{1}{2} \big(  \CM_3^{(a)}\CS^{(a)}(02) \otimes \CM_3^{(b)}\CA^{(b)}(02)
                               \nonumber \\
      & & +\CM_3^{(a)}\CA^{(a)}(02) \otimes \CM_3^{(b)}\CS^{(b)}(02) \Big).
\end{eqnarray}
The remaining elements are given as $E_2^{{\rm L}} = T(12)E_1^{{\rm L}}T(12)$ 
by the exchange symmetry, and $E_0^{{\rm L}} = 1-E_1^{{\rm L}}-E_2^{{\rm L}}$ by the 
completeness of the POVM.
We can now show that this separable POVM $E_\mu^{{\rm L}}$ can be implemented by a LOCC protocol, 
which is summarized as follows:
\begin{itemize}
\item First, Alice and Bob determine the permutation symmetry of their local system: totally symmetric, 
mixed symmetric, or totally antisymmetric. This is done by a projective measurement with 
the set of orthogonal projectors $\{\CS_3^{(p)}, \CM_3^{(p)}, \CA_3^{(p)}\}$ of each party $p(=a,b)$. 
Note that the cases $\CS_3^{(a)} \otimes \CA_3^{(b)}$ and $\CA_3^{(a)} \otimes \CS_3^{(b)}$ 
do not occur, since they would imply that the whole system is totally antisymmetric.
\item If their outcome is $\CS_3^{(a)} \otimes \CS_3^{(b)}$ or $\CA_3^{(a)} \otimes \CA_3^{(b)}$, 
Alice and Bob declare an inconclusive result, i.e., 0.
\item If one of the two parties $p(=a\ {\rm or }\ b)$ finds that his or her local system is 
totally symmetric, $\CS_3^{(p)}$, and the system of the other party $q(\ne p)$ 
is found to be mixed symmetric, $\CM_3^{(q)}$, then party $q$ performs a POVM measurement: 
\begin{eqnarray*}
  e_1 & \equiv & \frac{2}{3}\CM_3^{(q)}\CA^{(q)}(02),     \\ 
  e_2 & \equiv & \frac{2}{3}\CM_3^{(q)}\CA^{(q)}(01),     \\ 
  e_0 & \equiv & \frac{1}{3}\CM_3^{(q)}\left(1+2A^{(q)}\right).
\end{eqnarray*}
Note that the set $\{e_1,e_2,e_0\}$ is a POVM since $e_\mu \ge 0,\ (\mu=0,1,2)$ and 
$\sum_\mu e_\mu = \CM_3^{(q)}$. 
The final identification result by Alice and Bob is the measurement outcome 
$\mu(=0,1,2)$ of party $q$.  
\item If one of the two parties $p(=a\ {\rm or }\ b)$ finds that his or her local system is 
totally antisymmetric, $\CA_3^{(p)}$, and the system of the other party $q(\ne p)$ 
is found to be mixed symmetric, $\CM_3^{(q)}$, then party $q$ performs a POVM measurement: 
\begin{eqnarray*}
  e'_1 & \equiv & \frac{2}{3}\CM_3^{(q)}\CS^{(q)}(02),   \\ 
  e'_2 & \equiv & \frac{2}{3}\CM_3^{(q)}\CS^{(q)}(01),   \\ 
  e'_0 & \equiv & \frac{1}{3}\CM_3^{(q)}\left(1-2A^{(q)}\right).
\end{eqnarray*}
It is easily verified that the set $\{e'_1,e'_2,e'_0\}$ is also a POVM in $V_\CM^{(q)}$, and the 
final identification result of Alice and Bob is chosen to be the measurement outcome 
$\mu(=0,1,2)$ of party $q$. 
\item Finally, when the total system is found to be in $V_\CM^{(a)} \otimes V_\CM^{(b)}$, one of the two 
parties, say Alice, performs the following POVM measurement:
\begin{eqnarray*}
  && e_{11} \equiv \frac{1}{2}\CM_3^{(a)}\CA^{(a)}(02), \  
     e_{12} \equiv \frac{1}{2}\CM_3^{(a)}\CS^{(a)}(02), \\ 
  && e_{21} \equiv \frac{1}{2}\CM_3^{(a)}\CA^{(a)}(01), \
     e_{22} \equiv \frac{1}{2}\CM_3^{(a)}\CS^{(a)}(01).
\end{eqnarray*}
It is evident that the above set $\{e_{a_1a_2}\}_{a_1,a_2=1,2}$ forms a POVM in $V_\CM^{(a)}$. 
If Alice's outcome $a_1$ is equal to 1, Bob performs a projective measurement by 
the set of orthogonal projectors $\{f_b\}_{b=1,2}$:
\begin{eqnarray*}
   f_1 \equiv \CM_3^{(b)}\CS^{(b)}(02),\ 
   f_2 \equiv \CM_3^{(b)}\CA^{(b)}(02),
\end{eqnarray*}
otherwise, by the set of orthogonal projectors $\{f'_b\}_{b=1,2}$:
\begin{eqnarray*}
   f'_1 \equiv \CM_3^{(b)}\CS^{(b)}(01), \
   f'_2 \equiv \CM_3^{(b)}\CA^{(b)}(01).
\end{eqnarray*}
The final identification result $\mu$ of Alice and Bob is Alice's result $a_1$,  
if Alice's outcome $a_2$ coincides with Bob's outcome $b$. 
Otherwise, the final result is the inconclusive one, i.e., 0.
\end{itemize} 

Substituting explicit dimensions in Eq.(\ref{separable_trace_ES}), we obtain the optimal success 
probability with the LOCC protocol:   
\begin{eqnarray}
  p_{\max}^{{\rm L}} = \frac{1}{36d_ad_b(d_ad_b+1)} 
       \left( 11d_a^2d_b^2+d_a^2+d_b^2-13 \right ),
\end{eqnarray}
whereas the globally attainable success probability of Eq.(\ref{global_p}) in terms of dimensions 
$d_a$ and $d_b$ is given by  
\begin{eqnarray}
  p_{\max} = \frac{1}{3d_ad_b} (d_ad_b-1).
\end{eqnarray}
Although there is a finite gap between $p_{\max}$ and $p^{{L}}_{\max}$ as shown before, 
the numerical difference is not very large. 
For example, in the case of a two-qubit bipartite system ($d_a=d_b=2$), 
the optimal LOCC protocol gives $p_{\max}^{{\rm L}}=19/80$, whereas 
the globally attainable probability is given by $p_{\max}=1/4$. The difference is only 1/80. 
In the limit of $d_a$ and $d_b$ going to infinity, we find that $p_{\max}^{{\rm L}}$ approaches 
11/36 and $p_{\max}$ approaches 1/3 with the difference 1/36.

\vspace{1ex}
%%%%%%%%%%%%%%%%%%%%%%%%%%%%%%%%%%%%%%%%%%%%%%%%%%%%%%%%%%
%%%%%%%%%%%%%%%%%%%%%%%%%%%%%%%%%%%%%%%  Concluding Remarks
%%%%%%%%%%%%%%%%%%%%%%%%%%%%%%%%%%%%%%%%%%%%%%%%%%%%%%%%%%
\section{Concluding Remarks}
It is known that two bipartite pure states can be optimally discriminated 
within a LOCC scheme if classical knowledge of the two states are available. 
In this paper, we investigated the identification problem of two bipartite pure states, 
where no classical knowledge of the reference states is given but only a copy of each 
reference state is available. The two reference states are independently and randomly 
chosen from the state space in a unitary invariant way. 

In the case of minimum-error identification, we found that the optimal 
identification can be done locally. This is true for any prior probabilities, 
but we assumed that the number $N$ of copies of each state is 1.  
In the limit of large $N$, the identification problem reduces to the 
standard discrimination problem. This is because one can obtain complete 
classical information on the reference states by performing a tomographical 
measurement on infinitely many copies of them. We note that an infinite number 
of copies of a bipartite state is not equivalent to perfect classical knowledge 
of the state. In general, possession of an unlimited number of copies is more 
advantageous than mere classical knowledge of the state, because two parties 
sharing the states can utilize a quantum channel. 
However, for the identification problems with $N=\infty$ considered 
in this paper, it turns out not to be an advantage. This is because the discrimination 
of two known pure states can be performed optimally by LOCC and no scheme 
can perform better than the optimal global protocol. 
Therefore, we can say that the pure-state identification can be optimally 
performed by means of LOCC when $N=1$ and $N=\infty$. 
We conjecture that this is also true for arbitrary $N$. 

On the other hand, we have demonstrated that any LOCC scheme for the 
unambiguous identification of two bipartite pure states cannot attain 
the maximum success probability achieved by the global measurement at 
least for the case of equal prior probabilities. 
This contrasts remarkably with the results for the standard discrimination 
and the minimum-error identification. Our result provides an example of 
nonlocality in distinguishing two pure states.
It is an interesting problem to study the unambiguous identification of 
bipartite pure states when the number of copies of the reference states 
is finite but greater than 1. 

Just for completeness, we comment on the $N=0$ case, where 
no measurement (global or LOCC) on the input state can
improve the success probability. The maximum success probabilities 
are $\max \{\eta_1,\eta_2\}$ and 0 for the minimum-error and 
the unambiguous identification problems, respectively.

In this paper we assumed that the two reference states are independently and uniformly 
distributed in the state space. This is a sensible assumption to make for two completely 
unknown states. However, we note that the optimal identification protocol depends on 
the distribution of the reference states and whether the LOCC scheme attains global 
optimality may also change for different distributions.


\vfill
\begin{thebibliography}{}
\bibitem{Helstrom76}
C.~W.~Helstrom,
{\it Quantum Detection and Estimation Theory} 
(Academic Press, New York, 1976).

\bibitem{Holevo82}
A.~S.~Holevo,
{\it Probabilistic and Statistical Aspects of Quantum Theory} 
(North-Holland, Amsterdam, 1982).

\bibitem{Ivanovic87}
I.~D.~Ivanovic,
Phys. Lett. A {\bf 123}, 257 (1987).

\bibitem{Dieks88}
D.~Dieks,
Phys. Lett. A {\bf 126}, 303 (1988).

\bibitem{Peres88}
A.~Peres,
Phys. Lett. A {\bf 128}, 19 (1988).

\bibitem{Wootters82}
W.~K.~Wootters and W.~H.~Zurek,
Nature {\bf 299}, 802 (1982). 

\bibitem{Peres91}
A.~Peres and W.~K.~Wootters, 
Phys. Rev. Lett. {\bf 66}, 1119 (1991).

\bibitem{Ban97}
M.~Ban, K.~Kurokawa, R.~Momose, and O.~Hirota, 
Int. J. Theor. Phys. {\bf 36}, 1269 (1997).

\bibitem{Sasaki98}
M.~Sasaki, K.~Kato, M.~Izutsu, and O.~Hirota, 
Phys. Rev. A {\bf 58}, 146 (1998).

\bibitem{Eldar01}
Y.~C.~Eldar and G.~D.~Forney, Jr., 
IEEE Trans. Inf. Theory {\bf 47}, 858 (2001). 

\bibitem{Bennett99}
C.~H.~Bennett, D.~P.~DiVincenzo, C.~A.~Fuchs, T.~Mor, 
E.~Rains, P.~W.~Shor, J.~A.~Smolin, and W.~K.~Wootters, 
Phys. Rev. A{\bf 59}, 1070 (1999).

\bibitem{Koashi07}
Masato Koashi, Fumitaka Takenaga, Takashi Yamamoto, and Nobuyuki Imoto, 
arXiv:0709.3196v1 [quant-ph]. 

\bibitem{Walgate00}
Jonathan~Walgate, Anthony~J.~Short, Lucien~Hardy, and Vlatko~Vedral,
Phys. Rev. Lett. {\bf 85}, 4972 (2000).

\bibitem{Virmani01}
S.~Virmani, M.~F.~Sacchi, M.~B.~Plenio, and D.~Markham,
Phys. Lett. A{\bf 288}, 62 (2001).

\bibitem{Chen01}
Y.-X.~Chen and D.~Yang, 
Phys. Rev. A{\bf 64}, 064303 (2001)

\bibitem{Chen02}
Y.-X.~Chen and D.~Yang, 
Phys. Rev. A{\bf 65}, 022320 (2002)

\bibitem{Ji05}
Zhengfeng~Ji, Hongen~Cao, and Mingsheng~Ying,
Phys. Rev. A{\bf 71}, 032323 (2005). 

\bibitem{Hayashi05}
A.~Hayashi, M.~Horibe, and T.~Hashimoto, 
Phys. Rev. A{\bf 72}, 052306 (2005). 

\bibitem{Bergou05}
Janos~A.~Bergou and Mark~Hillery,
Phys. Rev. Lett. {\bf 94}, 160501 (2005).

\bibitem{Hayashi06}
A.~Hayashi, M.~Horibe, and T.~Hashimoto,  
Phys. Rev. A{\bf 73}, 012328 (2006). 

\bibitem{Bergou06}
Janos~A.~Bergou, Vladimir~Buzek, Edgar~Feldman, Ulrike~Herzog, and Mark~Hillery,
Phys. Rev. A{\bf 73}, 062334 (2006).

\bibitem{Hayashi04}
A.~Hayashi, T.~Hashimoto, and M.~Horibe, 
Phys. Rev. A{\bf 72}, 032325 (2005). 

\bibitem{Hamermesh62}
M.~Hamermesh,
{\it Group Theory and its Application to Physical Problems},
(Addison-Wesley, Reading, MA, 1962).

\end{thebibliography}
\end{document}